%% file: main.tex
\documentclass[10pt,twocolumn,a4paper, margin=1cm]{article}
\usepackage[a4paper, margin=2.5cm]{geometry}
\usepackage[pagebackref=true,breaklinks=true,colorlinks,bookmarks=false]{hyperref}
\usepackage[english]{babel}

\usepackage{graphicx}
\usepackage{amsmath}
\usepackage{amssymb}

\usepackage{listings}
\usepackage{xcolor}

\definecolor{mygreen}{rgb}{0,0.6,0}
\definecolor{light-gray}{rgb}{0.96,0.96,0.96}
\definecolor{mygray}{rgb}{0.5,0.5,0.5}
\definecolor{mymauve}{rgb}{0.58,0,0.82}

\newcommand{\lib}{\textit{TorchRadon}}
\input{cmds.tex}

\title{\lib: Fast Differentiable Routines for Computed Tomography}

\author{Matteo Ronchetti \\
Università di Pisa \\
{\tt\small mttronchetti@gmail.com}
}

\begin{document}

\maketitle

\begin{abstract}
   This work presents \lib{} -- an open source CUDA library which contains a set of differentiable routines for solving computed tomography (CT) reconstruction problems.
   The library is designed to help researchers working on CT problems to combine deep learning and model-based approaches.
   The package is developed as a \textit{PyTorch} extension and can be seamlessly integrated into existing deep learning training code.
   Compared to the existing Astra Toolbox, \lib{} is up to $125\times$ faster.
   The operators implemented by \lib{} allow the computation of gradients using PyTorch \textit{backward()}, and can therefore be easily inserted inside existing neural networks architectures.
   Because of its speed and GPU support, \lib{} can also be effectively used as a fast backend for the implementation of iterative algorithms.
   This paper presents the main functionalities of the library, compares results with existing libraries and provides examples of usage.
\end{abstract}

\vspace{0.5cm}
\section{Introduction}
In computed tomography (CT) the inner structure of a physical body is reconstructed from a series of external measurements.
Typically these consist of a series of X-ray images taken from different directions.
For example in diagnostic radiology a cross-section of the human body is scanned by a thin X-ray beam whose intensity loss is recorded by a detector and processed to produce a two-dimensional image.

The problem of CT reconstruction when a comprehensive, dense set of projections views are available is well studied. Classical methods (filtered backprojection in particular) are known to yield sub-optimal performance when dealing with limited, sparse or noisy tomographic data. \\
Regularization methods are usually adopted to tackle the ill-posedness of CT reconstructions \cite{ill_conditioned}, in particular total variation is frequently used, but also wavelets \cite{wavelets1}, curvelets \cite{curvelets1} and shearlets \cite{shearlets1, shearlets2, shearlets3} have  been  successfully  applied.

In recent years deep learning approaches have been used to tackle CT reconstruction problems yielding impressive results and often outperforming model-based approaches, which used to be the previous state of the art.
We refer the interested reader to the reviews \cite{review_1,review_2} for a detailed discussion on the use of deep neural networks in the context of inverse problems. \\
Although being highly successful current deep learning approaches require large amounts of training data, making them impractical to use for problems where data is scarce.
Furthermore the black-box nature of most neural networks could be a critical barrier for their application in the medical field. \\
Recent works use a combination of model-based and data-based (deep learning) approaches to overcome these limitations.
Deep neural networks have been inserted into iterative reconstruction schemes by unrolling the steps and casting them as a neural network \cite{lista, psidonet}.
Convolutional neural networks (CNNs) have been used to replace some proximal operators used by iterative reconstruction schemes \cite{proximal_op, proximal_op_2}.
A different approach is taken by \cite{Bubba_2019} which uses $\ell$1 shearlet regularization to decompose the reconstruction into \textit{visible} and \textit{invisible} coefficients and trains a CNN to predict the invisible coefficients.

The \lib{} library is designed to help researchers working on CT problems to combine deep learning and model-based approaches.
The library extends the PyTorch \cite{pytorch} deep learning library with routines specific to computed tomography and regularization of inverse problems. \\
Operations are implemented with optimized CUDA kernels allowing to fully utilize the computational power of modern GPUs and are integrated with PyTorch Autograd system. Therefore these routines can be used as layers within neural networks without requiring any change to training code. \\
Existing libraries (like Astra Toolbox \cite{astra:1, astra:2}) are not designed with this objective, their integration with deep learning frameworks can be problematic and their performance sub-optimal.

The source code of \lib{} is made publicly available on Github\footnote{\url{https://github.com/matteo-ronchetti/torch-radon}} under a free software license (GNU General Public License v3.0). Precompiled packages are made available for multiple versions of PyTorch, CUDA and Python\cite{python} under the Linux operating systems. \\
The library can be easily tried using online platforms like Google Colaboratory (see links and examples inside the Github repository).

\section{TorchRadon Overview}
The most important features of \lib{} are:
\begin{itemize}
    \item \textbf{Differentiability}: Having differentiable functions means that operators can be placed as layers within neural networks trained via backpropagation. In the presented library Radon forward and backward projections, and shearlet transforms are implemented as differentiable operators.
    \item \textbf{Speed}: \lib{} is up to $125\times$ faster than Astra Toolbox in computing Radon forward and Backward projections. The main reasons for this difference are that \lib{} works directly on data stored on the GPU and can fully utilize these processors by doing batch operations. See Section \ref{benchmarks} for a detailed performance comparison between \lib{} and other libraries.
    \item \textbf{Transparent API}: all the operations are seamlessly integrated with PyTorch \cite{pytorch}. Gradient can be computed using PyTorch \textit{backward()}, half precision can be used with Nvidia AMP \footnote{See \url{https://github.com/NVIDIA/apex}, now integrated in PyTorch \url{https://pytorch.org/docs/stable/notes/amp_examples.html}}.
    \item \textbf{Parallel programming}: batch processing allows to fully exploit the power of modern GPUs by processing multiple images in parallel. All the \lib{}'s functions, including solvers, support batch processing.
    \item \textbf{Half Precision}: Storing data in half precision (16bits) allows to get sensible speedups when doing Radon forward and backward projections with a very small accuracy loss. Refer to section \ref{half_precision} for more details about the loss of numerical accuracy.
\end{itemize}

\subsection{Package Structure}
The library is divided into submodules grouped by functionality:
\begin{itemize}
    \item \textbf{torch\_radon}: main module containing Radon projections. Currently parallel beam and fan-beam projections are implemented.
    \item \textbf{torch\_radon.shearlets}: GPU implementation of Alpha Shearlet Transform \footnote{Based on \url{https://github.com/dedale-fet/alpha-transform}}.
    Shearlets have been successfully applied to computed tomography problems \cite{shearlets1, shearlets2, Bubba_2019} and also to other inverse problems, such as denoising \cite{shearlet_denoising}, phase retrieval \cite{shearlet_phase} and inverse scattering \cite{shearlet_inverse_scattering}.
    \item \textbf{torch\_radon.solvers}: module containing common iterative algorithms for solving the tomography reconstruction problem. Currently Conjugate Gradient (CG), Conjugate Gradient on the Normal Equations (CGNE) \cite{cgne} and Landweber iteration \cite{landweber} are implemented.
\end{itemize}

\begin{figure}
    \centering
    \includegraphics[width=0.9\linewidth]{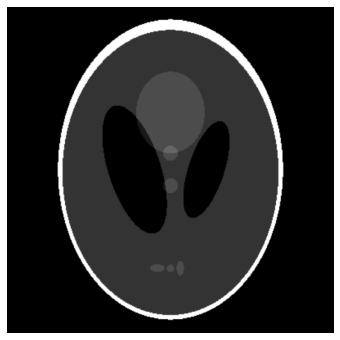}
    \caption{The classic Shepp-Logan phantom}
    \label{phantom}
\end{figure}

\section{Quality of Results}
In this section we compare the results obtained by \lib{} and other similar libraries. \\
We first compare the outputs of \lib{} and Astra Toolbox \cite{astra:1,astra:2} on Radon forward and backward projections and filtered backprojection. \\
Next, results of shearlet transforms are compared with results obtained by AlphaTransforms. \\
Finally, we describe the rationale behind the idea of storing inputs and outputs of Radon transforms in half-precision (16bits) while doing computations in single (32bits) precision and quantify the loss of accuracy incurred by using half precision.

The code snippets shown in this section assumes the following imports:
\begin{lstlisting}
import numpy as np
from torch_radon import Radon, RadonFanbeam
\end{lstlisting}

The code used for this comparison makes use of the Matplotlib library \cite{matplotlib} for visualization and can be found\footnote{Here \url{https://github.com/matteo-ronchetti/torch-radon/blob/master/examples/Figures\%20for\%20paper.ipynb}} in the examples folder on \lib{}'s git repository.
A more detailed comparison can be found in the unit tests\footnote{Located in the folder \url{https://github.com/matteo-ronchetti/torch-radon/tree/master/tests}} included in the source code. These tests are used to check the correctness of the library's implementation before each release.

\subsection{Comparison with Astra Toolbox}
We compare the results of Radon transforms and Filtered Bacprojetion (FBP) reconstruction against the ones obtained by Astra Toolbox. For the visual comparison we use the classical Shepp-Logan phantom (depicted in Figure \ref{phantom}) with size $512\times512$.

\subsubsection{Parallel Beam Projection}
\label{parallel_beam}
\begin{figure}[h]
\centering
   \includegraphics[width=0.95\linewidth]{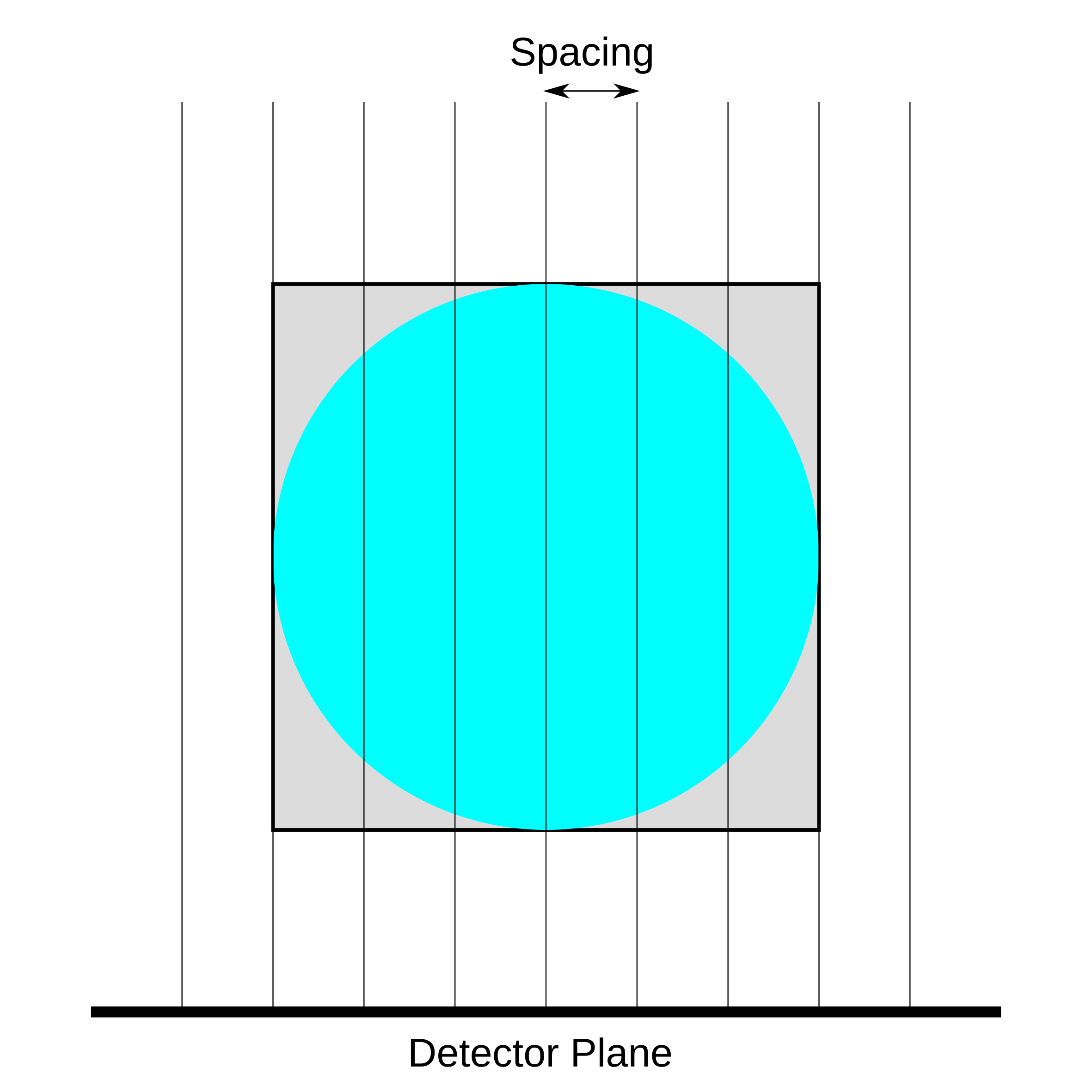}
\caption{Visualization of parallel beam projection.}
\label{fig:pb_geom}
\end{figure}

\begin{figure*}
\centering
   \includegraphics[width=0.95\linewidth]{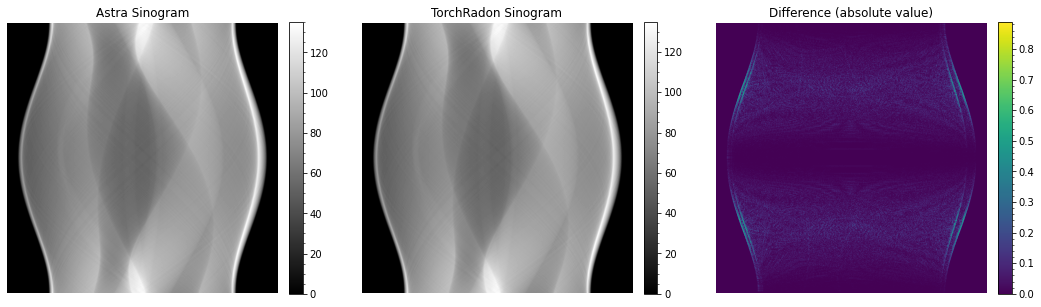}
\caption{Comparison of forward Radon transform with parallel beam projection.}
\label{fig:parallel_forward}
\end{figure*}
\begin{figure*}
\centering
   \includegraphics[width=0.95\linewidth]{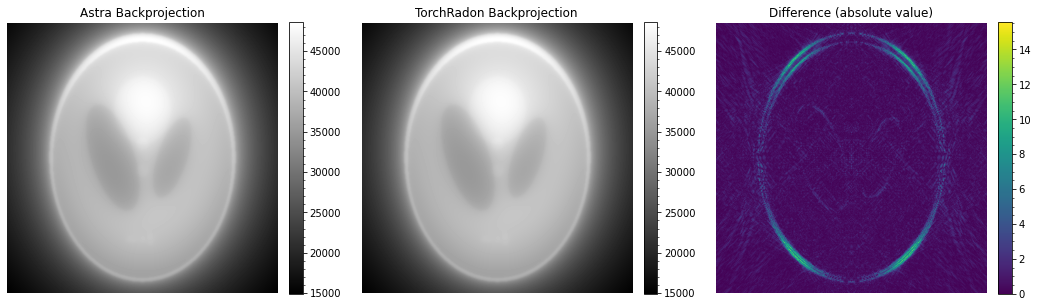}
\caption{Comparison of backward Radon transform with parallel beam projection.}
\label{fig:parallel_backward}
\end{figure*}

The geometry of parallel beam projection is summarized in Figure \ref{fig:pb_geom}.
For the comparison we use $512$ sampling angles equally spaced in $[0, \pi)$, the detector is given $512$ pixel with a spacing of $1.0$.
This setting corresponds to the following code in \lib{}:
\begin{lstlisting}
angles = np.linspace(0, np.pi, 512, endpoint=False)
radon = Radon(512, angles)

# Radon forward projection
sino = radon.forward(x)

# Radon backward projection
bp = radon.backprojection(sino)
\end{lstlisting}

Results of forward projection on the phantom are visualized in Figure \ref{fig:parallel_forward}. Let $y^*$ be the sinogram computed by Astra Toolbox and $y$ the sinogram computed by \lib{}, then the relative error $\frac{\norm{y^* - y}}{\norm{y}}$ is $8.07 \cdot 10^{-4}$.

Results of backward projection are depicted in Figure \ref{fig:parallel_backward}. For backprojection the relative error is $4.12 \cdot 10^{-5}$.

\subsubsection{Fan-Beam Projection}
\begin{figure}[h]
\centering
 \includegraphics[width=0.95\linewidth]{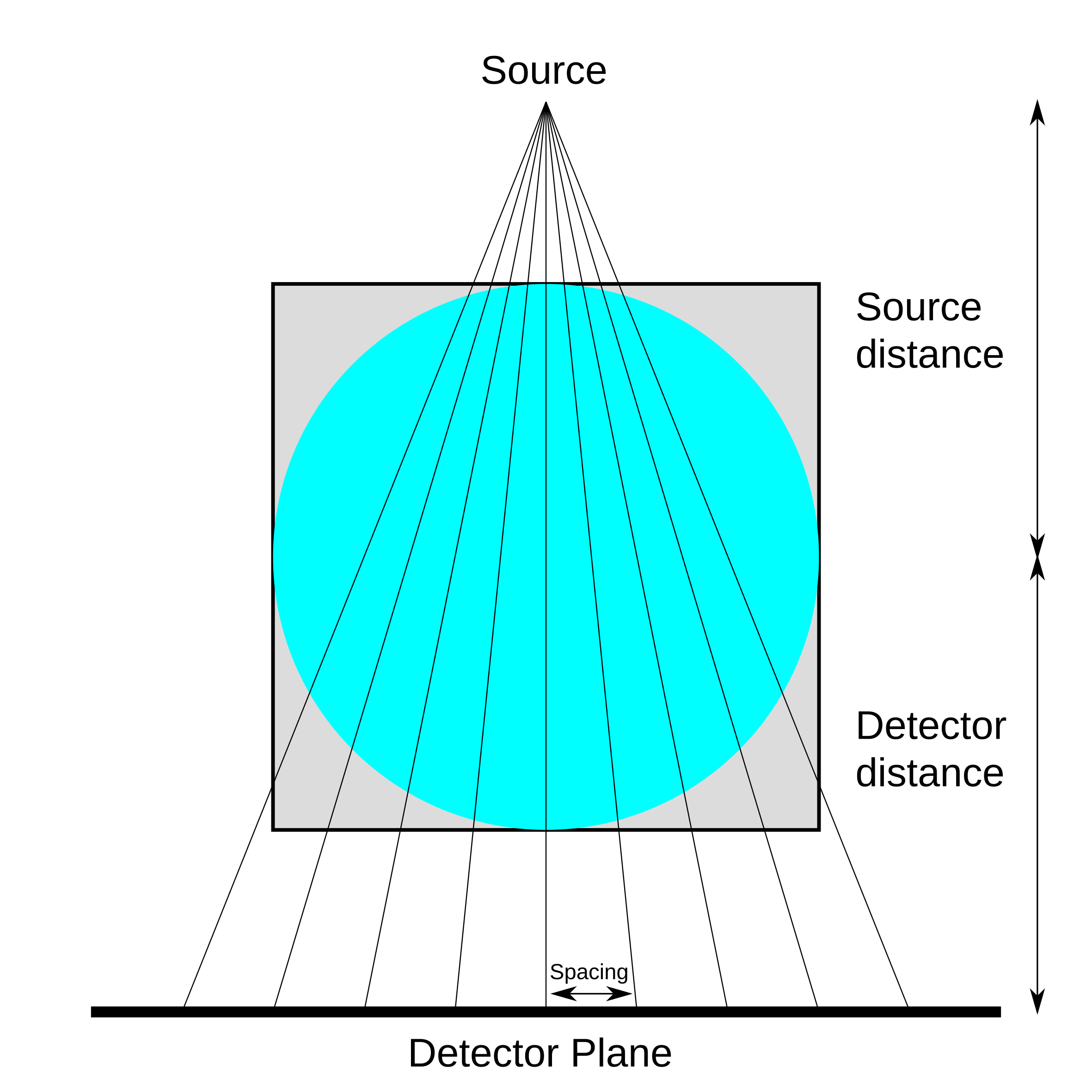}
\caption{Visualization of fan-beam projection.}
\label{fig:fb_geom}
\end{figure}
The geometry of parallel beam projection is summarized in Figure \ref{fig:fb_geom}. X-Rays are generated by a point source located at distance \textit{source\_distance} from the center of the object. The detector is a plane that contains \textit{det\_count} pixels contiguously spaced at distance \textit{det\_spacing}. The detector is located at distance \textit{det\_distance} from the center of the image.

For the comparison we use $512$ sampling angles equally spaced in $[0, 2\pi)$, a X-Ray source at distance $\textit{source\_distance} = 512$ and place the detector at the same distance ($\textit{det\_distance} = 512$). The detector is given $512$ pixels with spacing $2.0$ so that the rays will cover the whole image.
This setting corresponds to the following code in \lib{}:
\begin{lstlisting}
angles = np.linspace(0, 2*np.pi, 512, endpoint=False)
radon = RadonFanbeam(512, angles, source_distance=512)

# Radon forward projection
sino = radon.forward(x)

# Radon backward projection
bp = radon.backprojection(sino)
\end{lstlisting}
In the above code we made use of \lib{}'s \textit{RadonFanbeam} default values: \textit{det\_distance} by default is assumed equal to  \textit{source\_distance} and \textit{det\_spacing} when not specified is the minimum value such that the projected rays will cover the whole image.

\begin{figure*}
\centering
   \includegraphics[width=0.95\linewidth]{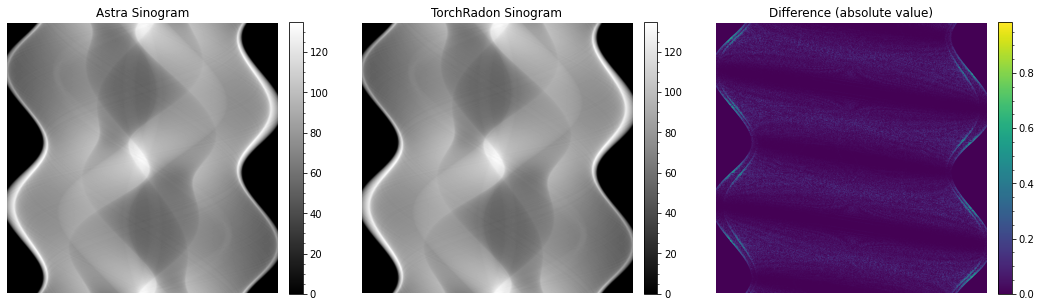}
\caption{Comparison of forward Radon transform with fan-beam projection.}
\label{fig:fanbeam_forward}
\end{figure*}
\begin{figure*}
\centering
   \includegraphics[width=0.95\linewidth]{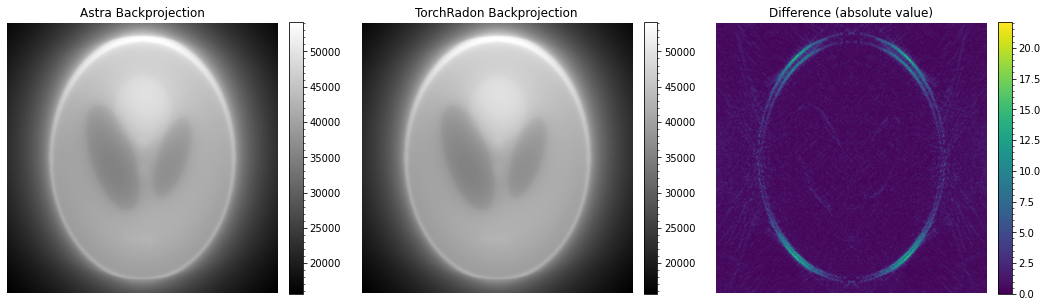}
\caption{Comparison of backward Radon transform with fan-beam projection.}
\label{fig:fanbeam_backward}
\end{figure*}

Results of forward projection on the phantom are visualized in Figure \ref{fig:fanbeam_forward}. Let $y^*$ be the sinogram computed by Astra Toolbox and $y$ the sinogram computed by \lib{}, then the relative error $\frac{\norm{y^* - y}}{\norm{y}}$ is $8.34 \cdot 10^{-4}$.

Results of backward projection are depicted in Figure \ref{fig:fanbeam_backward}. For backprojection the relative error is $4.84 \cdot 10^{-5}$.

\subsubsection{Filtered Backprojection}
\label{fbp}
\begin{figure*}
\centering
   \includegraphics[width=0.95\linewidth]{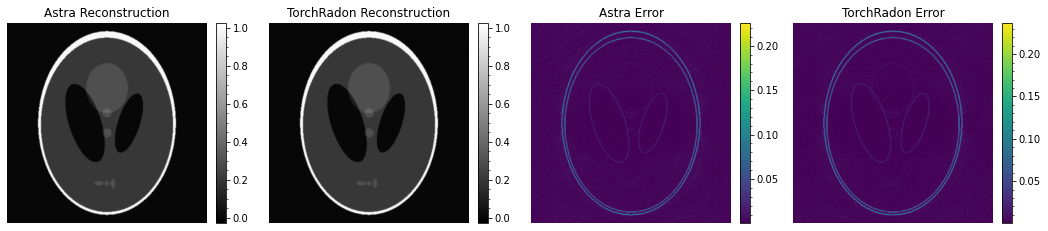}
\caption{Comparison of reconstructions using FBP with Ram-Lak filter.}
\label{fig:fbp}
\end{figure*}

Filtered backprojection (FBP) can be used to reconstruct the original image $x$ given fully sampled and noiseless measurements $y$. \lib{} implements the filtration of the sinogram $y$ in the frequency domain. Currently the library includes the following filters: Ram-Lak (ramp), Shepp-Logan, cosine, Hamming, Hann. To reduce artifacts the construction of the Fourier filter is done as explained in \cite{fbp_filtering}, Chap 3. Equation 61.

We use parallel beam projection with $512$ sampling angles equally spaced in $[0, \pi)$. The detector is given $\left\lceil 512 \, \sqrt{2} \, \right\rceil$ pixels with spacing of $1.0$ so that reconstruction could be exact also for pixels that lies outside of the inscribed circle. Ram-Lak filter is used to filter the sinogram. This setting corresponds to the following code in \lib{}:
\begin{lstlisting}
angles = np.linspace(0, np.pi, 512, endpoint=False)
det_count = int(np.ceil(np.sqrt(2)*512))

radon = Radon(512, angles, det_count=det_count)

sino = radon.forward(x)
filtered_sino = radon.filter_sinogram(sino, "ram-lak")
fbp = radon.backprojection(filtered_sino)
\end{lstlisting}
Figure \ref{fig:fbp} depicts the reconstruction results. Astra Toolbox achieves a Mean Squared Error (MSE) of $2.02 \cdot 10^{-4}$ while \lib{} obtains a MSE of $2.22 \cdot 10^{-4}$.

\subsection{Comparison with AlphaTransforms}
Our implementation of the Alpha Shearlet transform is based on the AlphaTransforms library\footnote{\url{https://github.com/dedale-fet/alpha-transform}}.
Fourier coefficients are computed once at initialization (can be cached on disk) by the AlphaTransforms library which is specified as a dependency.
The main difference with the AlphaTransforms library is that, once the Fourier coefficients are computed, all the subsequent operations are done entirely on the GPU and can be used to process a batch images in parallel.

We compare the results of \lib{} and the AlphaTransforms library using real-valued alpha-shearlets with $\alpha = 0.5$ and $5$ scales. Alpha-shearlets are normalized (on the Fourier side) to get a Parseval frame, therefore the composition of the transform with its adjoint is the identity mapping. \\
This setting corresponds to the following code in \lib{}:
\begin{lstlisting}
n_scales = 5

shearlet = ShearletTransform(512, 512, [0.5] * n_scales)

# compute shearlet coefficients
coeff = shearlet.forward(x)

# adjoint transform reconstructs the image
rec = shearlet.backward(coeff)
\end{lstlisting}

\begin{figure*}
\centering
   \includegraphics[width=0.95\linewidth]{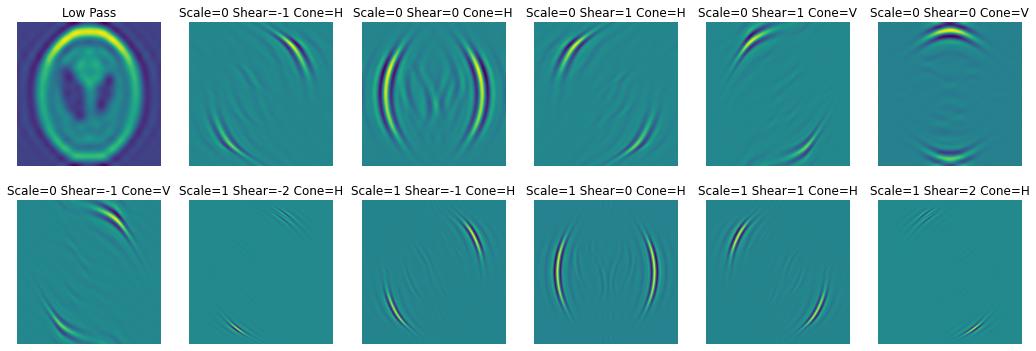}
\caption{12 coefficients of the shearlet transform of the Shepp-Logan phantom.}
\label{fig:shear_coeff}
\end{figure*}

\begin{figure*}
\centering
   \includegraphics[width=0.95\linewidth]{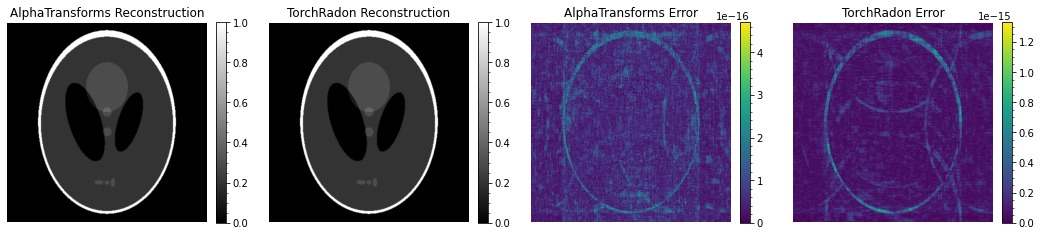}
\caption{Shearlet reconstructions obtained by AlphaTransforms and \lib{} together with   the absolute value of the pixel-wise error}
\label{fig:shear_rec}
\end{figure*}

This setting produces $59$ shearlet coefficients, 12 of which are depicted in Figure \ref{fig:shear_coeff}. \\
The relative error between AlphaTransforms' coefficients and \lib{}'s ones is $3.86 \cdot 10^{-7}$ when using \lib{} with single precision (32bits), and $4.74 \cdot 10^{-16}$ when using double precision (64bits). \\
Reconstructions obtained by both the libraries are shown in Figure \ref{fig:shear_rec} together with the absolute value of the pixel-wise error.
The relative reconstruction error obtained by AlphaTransforms is $3.51 \cdot 10^{-16}$, while the one obtained by \lib{} is $6.14 \cdot 10^{-16}$ ($6.45 \cdot 10^{-7}$ using single precision).

\subsection{Half precision Radon transforms}
\label{half_precision}

\begin{figure*}
\centering
   \includegraphics[width=0.95\linewidth]{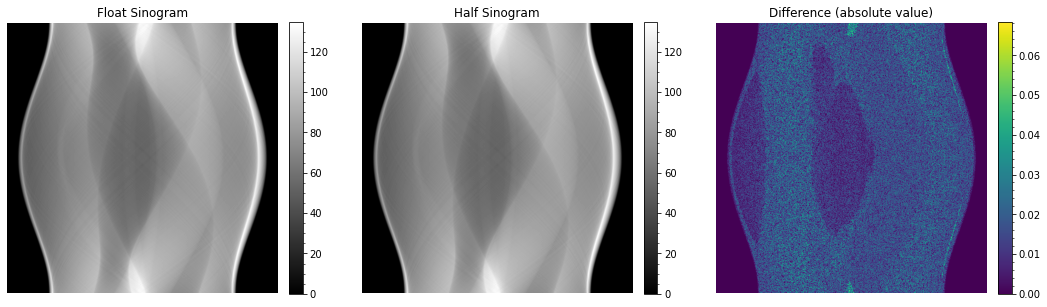}
\caption{Comparison between sinograms obtained using single precision (32bits) and half precision (16bits).}
\label{fig:hp_sino}
\end{figure*}

\begin{figure*}
\centering
   \includegraphics[width=0.95\linewidth]{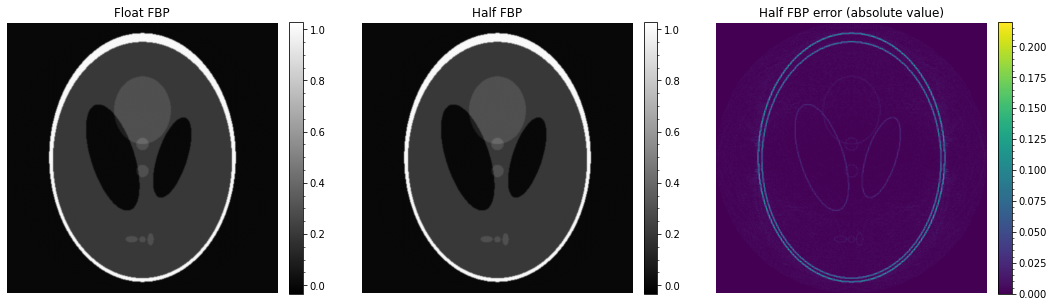}
\caption{Comparison between FBP reconstructions obtained using single precision (32bits) and half precision (16bits) data.}
\label{fig:hp_fbp}
\end{figure*}

Radon transforms are memory bound operations, the device spends most of the execution time doing memory reads while arithmetic operations only take a negligible amount of time. Furthermore the nature of Radon transforms makes it hard to get good cache hit rates.
\lib{} stores input data in CUDA Textures which are read-only data structures that use opaque memory layouts optimized for texture fetching. Texture reads are cached by the GPU in a two-dimensional neighborhood, improving cache hit rates for operations like line integrals required by Radon transforms.

When given half precision (16bits) data, \lib{} packs 4 images into the channels of a single texture therefore reducing the number of memory reads by a factor of 4. Once read the data is converted to single precision (32bits) and all the arithmetic operations are done in single precision to minimize the loss of accuracy. \\
The speedup with respect to single precision is more than $2.5\times$, refer to Section \ref{batch} and Figure \ref{fig:radon_benchmark} for further details about performances.

We quantify the loss of accuracy by comparing the results obtained using single and half precision. \\
Let $x$ be the Shepp-Logan phantom in single precision and $\widetilde{x}$ be the same phantom stored in half precision, then the relative error $\frac{\norm{x - \widetilde{x}}}{\norm{x}}$ between single and half representation is $1.40 \cdot 10^{-4}$. \\
The sinograms obtained by Radon forward projection are depicted in Figure \ref{fig:hp_sino}, the relative error is $2.43 \cdot 10^{-4}$. \\
Figure \ref{fig:hp_fbp} shows the reconstructions obtained using Filtered Backprojection with single and half precision. The filtration of FBP in both cases is done in single precision. The mean squared reconstruction error with respect to the original phantom are almost equal being $2.0566\cdot 10^{-4}$ in single precision and $2.0567\cdot 10^{-4}$ when using half precision.

\section{Use Cases and Benchmarks}
\label{benchmarks}

This section presents practical usage examples of the library, demonstrating its ease of use, speed and capabilities. \\
First we show how to process multiple images in parallel using batch operations to fully exploit the computational power of modern GPUs. Compared to Astra Toolbox we obtain a speedup of more than $40\times$ ($125\times$ using half precision). \\
Next, we demonstrate the use of \lib{}'s solvers, reconstructing an image using Landweber iteration and CGNE. \\
Finally, we use \lib{}'s conjugate gradient solver and shearlet transform to implement the algorithm
(A.3) of \cite{Bubba_2019} and reducing its runtime by 125$\times$ with respect to the time reported  in the original paper.

For further details about the usage of the library please refer to the documentation\footnote{ \url{https://torch-radon.readthedocs.io}} and the examples provided together with the source code.

\subsection{Batch Image Processing}
\label{batch}
Batch processing allows to fully exploit the power of modern GPUs by processing multiple images in parallel. All \lib{}'s functions can take batch of data as inputs.
For example the code of Section \ref{parallel_beam} can be adapted to work with a batch of data just by changing the shape of $x$:
\begin{lstlisting}
# process 32 images in parallel
batch_size = 32

# create random images
x = torch.randn(batch_size, 512, 512).cuda()

angles = np.linspace(0, np.pi, 512, endpoint=False)
radon = Radon(512, angles)

# Radon forward projection
sino = radon.forward(x)

# Radon backward projection
bp = radon.backprojection(sino)
\end{lstlisting}

We compare the speed of \lib{} and Astra Toolbox when computing Radon forward and backward projections for both parallel beam and fan beam projections. \\
When using a GPU to train a neural network that contains Radon transforms, both the inputs and the outputs of the transforms need to be in GPU memory.
To emulate this situation in our benchmark, input data is a batch of multiple images (or sinograms) stored continuously in GPU memory. Similarly we force the output to be a continuous array on GPU, moving to the device it if necessary.

\begin{figure*}
\centering
   \includegraphics[width=0.45\linewidth]{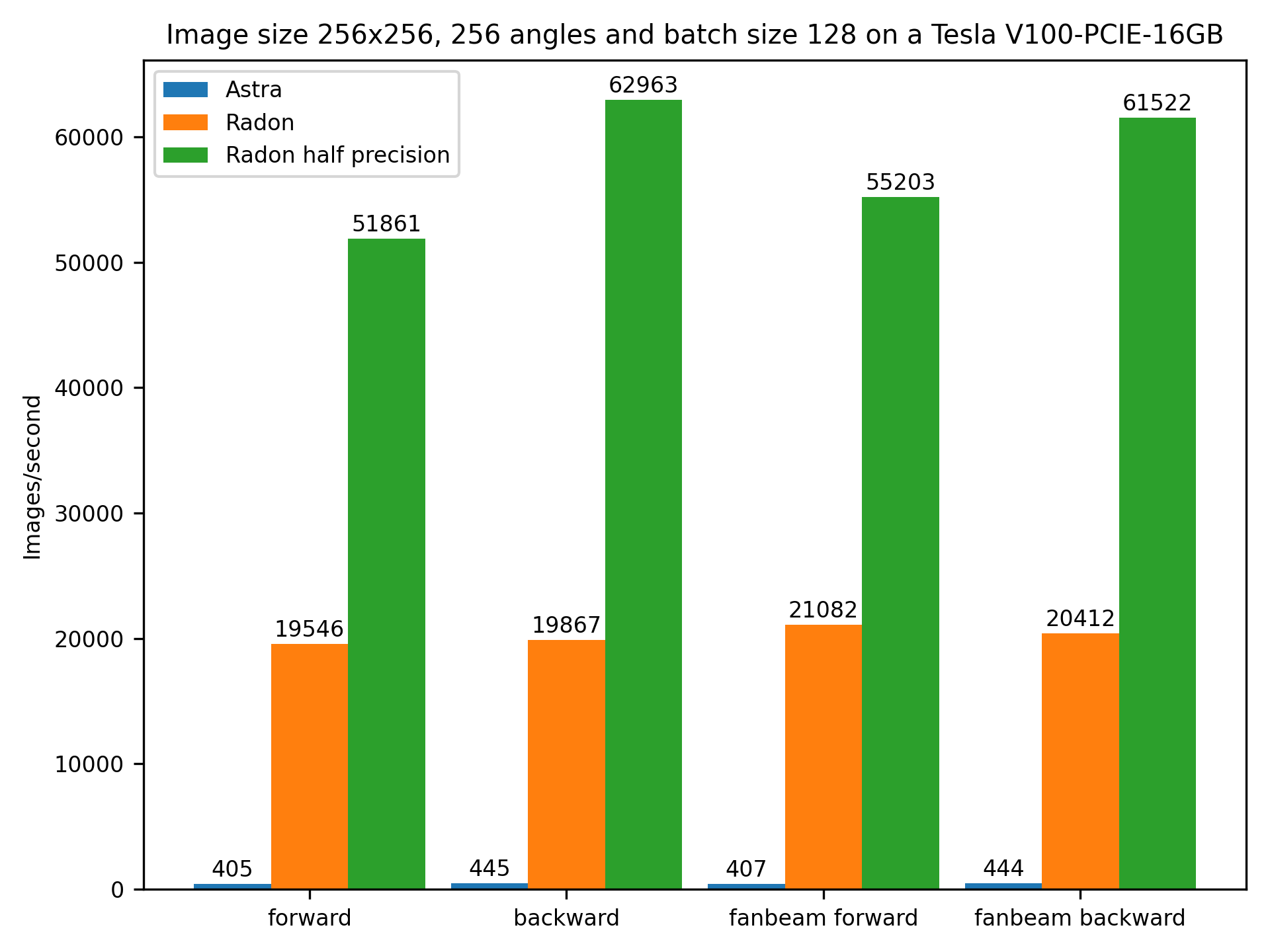}
\includegraphics[width=0.45\linewidth]{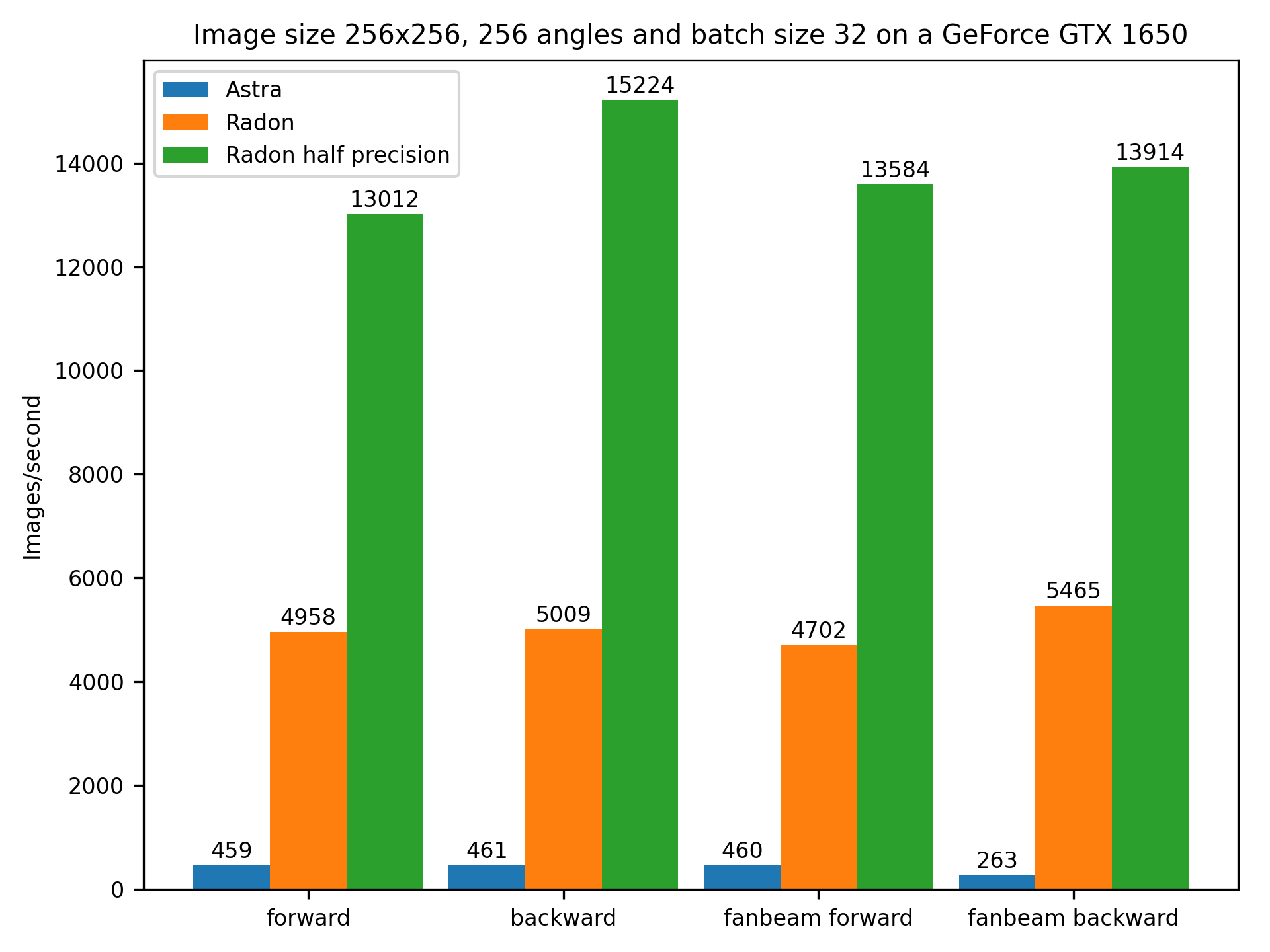}
\caption{Comparison of the speed of Torch Radon and Astra Toolbox on a modern server GPU (Tesla V100, left) and a laptop GPU (GeForce GTX 1650, right).}
\label{fig:radon_benchmark}
\end{figure*}

Results for both a modern server GPU (Tesla V100) and a laptop GPU (GTX 1650) are visualized in Figure \ref{fig:radon_benchmark}. When using single precision data \lib{} is more than $40\times$ faster than Astra Toolbox when running on a Tesla V100.
Furthermore, by making use of half precision (16bits) storage we are able to obtain speedups of more than $125\times$.

\subsection{Reconstruction with Iterative Solvers}
In this section we compare the reconstructions obtained using FBP, Conjugate Gradient on the Normal Equations (CGNE) \cite{cgne} and Landweber iteration \cite{landweber}. \\
We use the same setting as in Section \ref{fbp} and make use of the solvers included in \lib{}.

\begin{lstlisting}
from torch_radon.solvers import cgne, Landweber

angles = np.linspace(0, np.pi, 512, endpoint=False)
det_count = int(np.ceil(np.sqrt(2)*512))

radon = Radon(512, angles, det_count=det_count)
sino = radon.forward(x)

# FBP
filtered_sino = radon.filter_sinogram(sino, "ram-lak")
fbp = radon.backprojection(filtered_sino)

# Landweber
landweber = Landweber(radon)
# start with a solution guess which is all zeros
guess = torch.zeros(x.size(), device=x.device)
# estimate the step size using power iteration
alpha = landweber.estimate_alpha(512, device) * 0.95
landweber_rec = landweber.run(guess, sino, alpha, iterations=500)

# CGNE
guess = torch.zeros(x.size(), device=x.device)
cgne_rec = cgne(radon, guess, sino, max_iter=500)
\end{lstlisting}

FBP achieves a Mean Squared Error (MSE) of $2.22 \cdot 10^{-4}$, Landweber of $1.39 \cdot 10^{-4}$ and CGNE of $4.42 \cdot 10^{-5}$.

\subsection{Reconstruction with $\ell^1$ Shearlet Regularization}
Bubba et al. \cite{Bubba_2019} solves the following $\ell^1$ shearlet regularized CT reconstruction problem as a preprocessing step:
$$
\argmin_{f \geq 0} \norm{SH(f)}_{1,w} + \frac{1}{2} \norm{\mathcal{R} f - y}_2^2
$$
where $SH$ is the shearlet transform, $\mathcal{R}$ is the forward Radon projection (with limited angles) and $y$ contains the measured sinogram.
This minimization problem is solved using the alternating direction method of multipliers (ADMM) \cite{admm}, refer to Algorithm (A.3) of the paper for further details.

The implementation of (A.3) with \lib{} is quite simple, we solve the linear system using \lib{} conjugate gradient solver, use Radon projection and shearlet transform and implement the remaining operations using standard PyTorch functions. The algorithm implementation follows (A.3) and uses the same hyper-parameters. We report here the code of the implementation:
\begin{lstlisting}
from torch_radon.solvers import cg

def shrink(a, b):
    return (torch.abs(a) - b).clamp_min(0) * torch.sign(a)

n_scales = 5
angles = (np.linspace(0., 100., n_angles, endpoint=False)-50.0) / 180.0 * np.pi

radon = Radon(512, angles)
shearlet = ShearletTransform(512, 512, [0.5] * n_scales)

sinogram = radon.forward(x)
bp = radon.backward(sinogram)
sc = shearlet.forward(bp)

p_0 = 0.02
p_1 = 0.1
w = 3 ** shearlet.scales / 400
w = w.view(1, -1, 1, 1).cuda()

u_2 = torch.zeros_like(bp)
z_2 = torch.zeros_like(bp)
u_1 = torch.zeros_like(sc)
z_1 = torch.zeros_like(sc)
f = torch.zeros_like(bp)

for i in range(num_iterations):
    cg_y = p_0 * bp + p_1 * shearlet.backward(z_1 - u_1) + (z_2 - u_2)
    f = cg(lambda x: p_0 * radon.backward(radon.forward(x)) + (1 + p_1) * x, f.clone(), cg_y, max_iter=50)
    sh_f = shearlet.forward(f)

    z_1 = shrink(sh_f + u_1, p_0 / p_1 * w)
    z_2 = (f + u_2).clamp_min(0)
    u_1 = u_1 + sh_f - z_1
    u_2 = u_2 + f - z_2
\end{lstlisting}

Compared to an implementation made using Astra and AlphaTransforms  using \lib{} has the following advantages:
\begin{enumerate}
    \item All the operations, including shearlet transforms, are done on the GPU maximizing execution speed.
    \item There are are no CPU-GPU memory copies inside the main loop of the algorithm.
    \item It is possible to process multiple sinograms in parallel by making use of \lib{} batch processing capabilities without any change to the algorithm.
\end{enumerate}

\begin{figure*}
\centering
   \includegraphics[width=0.95\linewidth]{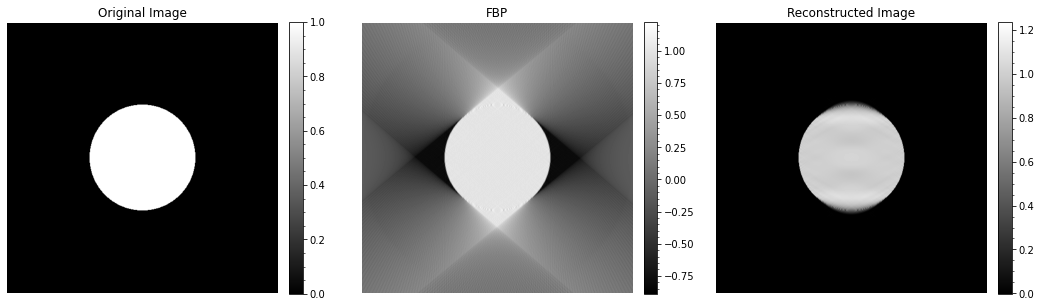}
\caption{Comparison between the reconstructions obtained using Filtered Backprojection and the ADMM method used in \cite{Bubba_2019}.}
\label{fig:admm}
\end{figure*}

To check the correctness of the implementation, results have been compared with the ones obtained by a Python implementation kindly shared by the authors of \cite{Bubba_2019}.
Figure \ref{fig:admm} shows the results of the reconstruction algorithm compared to FBP. \\
On a Tesla V100 GPU our implementation finishes in $1.6$ seconds, this is much faster than the $2.5$ minutes runtime reported in the original paper (on an Intel i7 CPU). Furthermore by increasing the batch size we can process multiple images in parallel obtaining an average time of $1.2$ seconds/image. Our straightforward implementation is therefore $125\times$ faster than the one used in \cite{Bubba_2019}.

\section{Conclusions}
We have introduced \lib{} an open source CUDA library which contains a set of differentiable routines for solving computed tomography reconstruction problems.
By making use of optimized GPU kernels, batch processing and by avoiding CPU-GPU copies, the presented library is up to two orders of magnitude faster than existing ones. \\
The integration with the PyTorch framework allows to easily integrate CT specific operations into classic neural networks architectures without any change to the training code.
We believe that the combination of speed and differentiability offered by \lib{} will be a key element in enabling the combination of deep learning and model-based approaches for CT reconstruction.

\bibliographystyle{amsplain}
\bibliography{bibliography}

\end{document}

%% file: cmds.tex
\usepackage[utf8]{inputenc}

\usepackage{amsfonts}
\usepackage{amsmath}
\usepackage{geometry}

\usepackage[parfill]{parskip}

\newcommand{\norm}[1]{\left\lVert#1\right\rVert}

\newcommand\opn{\mathrel{\ooalign{$\subseteq$\cr
  \hidewidth\raise.225ex\hbox{$\circ\mkern.5mu$}\cr}}}

\DeclareMathOperator*{\argmin}{arg\,min}

\usepackage{xcolor}

%% file: main.bbl
\providecommand{\bysame}{\leavevmode\hbox to3em{\hrulefill}\thinspace}
\providecommand{\MR}{\relax\ifhmode\unskip\space\fi MR }
\providecommand{\MRhref}[2]{%
  \href{http://www.ams.org/mathscinet-getitem?mr=#1}{#2}
}
\providecommand{\href}[2]{#2}
\begin{thebibliography}{10}

\bibitem{review_1}
Simon Arridge, Peter Maass, Ozan Öktem, and Carola-Bibiane Schönlieb,
  \emph{Solving inverse problems using data-driven models}, Acta Numerica
  \textbf{28} (2019), 1–174.

\bibitem{shearlets3}
T.~A. Bubba, M.~März, Z.~Purisha, M.~Lassas, and S.~Siltanen,
  \emph{{Shearlet-based regularization in sparse dynamic tomography}}, Wavelets
  and Sparsity XVII (Yue~M. Lu, Dimitri Van~De Ville, and Manos Papadakis,
  eds.), vol. 10394, International Society for Optics and Photonics, SPIE,
  2017, pp.~236 -- 245.

\bibitem{psidonet}
Tatiana~A Bubba, Mathilde Galinier, Matti Lassas, Marco Prato, Luca Ratti, and
  Samuli Siltanen, \emph{Deep neural networks for inverse problems with
  pseudodifferential operators: an application to limited-angle tomography},
  arXiv preprint arXiv:2006.01620 (2020).

\bibitem{Bubba_2019}
Tatiana~A Bubba, Gitta Kutyniok, Matti Lassas, Maximilian März, Wojciech
  Samek, Samuli Siltanen, and Vignesh Srinivasan, \emph{Learning the invisible:
  A hybrid deep learning-shearlet framework for limited angle computed
  tomography},  \textbf{35} (2019), no.~6, 064002.

\bibitem{shearlets2}
Tatiana~A. Bubba, Federica Porta, Gaetano Zanghirati, and Silvia Bonettini,
  \emph{A nonsmooth regularization approach based on shearlets for poisson
  noise removal in roi tomography}, Applied Mathematics and Computation
  \textbf{318} (2018), 131 -- 152, Recent Trends in Numerical Computations:
  Theory and Algorithms.

\bibitem{curvelets1}
Emmanuel~J. Candes and David~L. Donoho, \emph{{Curvelets and reconstruction of
  images from noisy radon data}}, Wavelet Applications in Signal and Image
  Processing VIII (Akram Aldroubi, Andrew~F. Laine, and Michael~A. Unser,
  eds.), vol. 4119, International Society for Optics and Photonics, SPIE, 2000,
  pp.~108 -- 117.

\bibitem{shearlets1}
Flavia Colonna, Glenn Easley, Kanghui Guo, and Demetrio Labate, \emph{Radon
  transform inversion using the shearlet representation}, Applied and
  Computational Harmonic Analysis \textbf{29} (2010), no.~2, 232 -- 250.

\bibitem{ill_conditioned}
Mark~E. Davison, \emph{The ill-conditioned nature of the limited angle
  tomography problem}, SIAM Journal on Applied Mathematics \textbf{43} (1983),
  no.~2, 428--448.

\bibitem{admm}
Jim Douglas, \emph{Alternating direction methods for three space variables},
  Numer. Math. \textbf{4} (1962), no.~1, 41–63.

\bibitem{shearlet_denoising}
G.~R. {Easley}, D.~{Labate}, and F.~{Colonna}, \emph{Shearlet-based total
  variation diffusion for denoising}, IEEE Transactions on Image Processing
  \textbf{18} (2009), no.~2, 260--268.

\bibitem{lista}
Karol Gregor and Yann LeCun, \emph{Learning fast approximations of sparse
  coding}, Proceedings of the 27th International Conference on International
  Conference on Machine Learning (Madison, WI, USA), ICML'10, Omnipress, 2010,
  p.~399–406.

\bibitem{cgne}
M.~Hanke, \emph{Conjugate gradient type methods for ill-posed problems.}, New
  York: Chapman and Hall/CRC.

\bibitem{matplotlib}
John~D Hunter, \emph{Matplotlib: A 2d graphics environment}, Computing in
  science \& engineering \textbf{9} (2007), no.~3, 90.

\bibitem{fbp_filtering}
A.~C. Kak and Malcolm Slaney, \emph{Principles of computerized tomographic
  imaging}, IEEE Press, 1998.

\bibitem{shearlet_inverse_scattering}
Gitta Kutyniok, Volker Mehrmann, and Philipp~C. Petersen, \emph{Regularization
  and numerical solution of the inverse scattering problem using shearlet
  frames}, Journal of Inverse and Ill-posed Problems \textbf{25} (01 Jun.
  2017), no.~3, 287 -- 309.

\bibitem{landweber}
L.~Landweber, \emph{An iteration formula for fredholm integral equations of the
  first kind}, American Journal of Mathematics \textbf{73} (1951), no.~3,
  615--624.

\bibitem{shearlet_phase}
Stefan Loock and Gerlind Plonka, \emph{Phase retrieval for fresnel measurements
  using a shearlet sparsity constraint}, Inverse Problems \textbf{30} (2014),
  no.~5, 055005.

\bibitem{wavelets1}
Ignace Loris, Guust Nolet, Ingrid Daubechies, and F.~A. Dahlen,
  \emph{{Tomographic inversion using $\ell$1-norm regularization of wavelet
  coefficients}}, Geophysical Journal International \textbf{170} (2007), no.~1,
  359--370.

\bibitem{review_2}
M.~T. {McCann}, K.~H. {Jin}, and M.~{Unser}, \emph{Convolutional neural
  networks for inverse problems in imaging: A review}, IEEE Signal Processing
  Magazine \textbf{34} (2017), no.~6, 85--95.

\bibitem{proximal_op}
Tim Meinhardt, Michael Moller, Caner Hazirbas, and Daniel Cremers,
  \emph{Learning proximal operators: Using denoising networks for regularizing
  inverse imaging problems}, Proceedings of the IEEE International Conference
  on Computer Vision, 2017, pp.~1781--1790.

\bibitem{pytorch}
Adam Paszke, Sam Gross, Soumith Chintala, Gregory Chanan, Edward Yang, Zachary
  DeVito, Zeming Lin, Alban Desmaison, Luca Antiga, and Adam Lerer,
  \emph{Automatic differentiation in {PyTorch}}, NIPS Autodiff Workshop, 2017.

\bibitem{python}
{Python Core Team}, \emph{{Python: A dynamic, open source programming
  language}}, {Python Software Foundation}, 2019.

\bibitem{astra:1}
Wim van Aarle, Willem~Jan Palenstijn, Jeroen Cant, Eline Janssens, Folkert
  Bleichrodt, Andrei Dabravolski, Jan~De Beenhouwer, K.~Joost Batenburg, and
  Jan Sijbers, \emph{Fast and flexible x-ray tomography using the astra
  toolbox}, Opt. Express \textbf{24} (2016), no.~22, 25129--25147.

\bibitem{astra:2}
Wim {van Aarle}, Willem~Jan Palenstijn, Jan {De Beenhouwer}, Thomas Altantzis,
  Sara Bals, K.~Joost Batenburg, and Jan Sijbers, \emph{The astra toolbox: A
  platform for advanced algorithm development in electron tomography},
  Ultramicroscopy \textbf{157} (2015), 35 -- 47.

\bibitem{proximal_op_2}
Yan Yang, Jian Sun, Huibin Li, and Zongben Xu, \emph{Deep admm-net for
  compressive sensing mri}, Proceedings of the 30th International Conference on
  Neural Information Processing Systems (Red Hook, NY, USA), NIPS'16, Curran
  Associates Inc., 2016, p.~10–18.

\end{thebibliography}
